\documentstyle[sprocl,epsfig]{article}

\begin{document}

\title{COMPARISON of WARM and COLD PHOTON COLLIDERS\\}

\author{ V.I.~TELNOV}

\address{Institute of Nuclear Physics,  630090, Novosibirsk, Russia}


\newcommand{\M}{\mbox{m}}
\newcommand{\n}{\mbox{$n_f$}}
\newcommand{\EP}{\mbox{e$^+$}}
\newcommand{\EM}{\mbox{e$^-$}}
\newcommand{\EPEM}{\mbox{e$^+$e$^-$}}
\newcommand{\EMEM}{\mbox{e$^-$e$^-$}}
\newcommand{\GG}{\mbox{$\gamma\gamma$}}
\newcommand{\GE}{\mbox{$\gamma$e}}
\newcommand{\GP}{\mbox{$\gamma$e$^+$}}
\newcommand{\TEV}{\mbox{TeV}}
\newcommand{\GEV}{\mbox{GeV}}
\newcommand{\LGG}{\mbox{$L_{\gamma\gamma}$}}
\newcommand{\LGE}{\mbox{$L_{\gamma e}$}}
\newcommand{\LEE}{\mbox{$L_{ee}$}}
\newcommand{\WGG}{\mbox{$W_{\gamma\gamma}$}}
\newcommand{\EV}{\mbox{eV}}
\newcommand{\CM}{\mbox{cm}}
\newcommand{\MM}{\mbox{mm}}
\newcommand{\NM}{\mbox{nm}}
\newcommand{\MKM}{\mbox{$\mu$m}}
\newcommand{\SEC}{\mbox{s}}
\newcommand{\CMS}{\mbox{cm$^{-2}$s$^{-1}$}}
\newcommand{\MRAD}{\mbox{mrad}}
\newcommand{\IND}{\hspace*{\parindent}}
\newcommand{\E}{\mbox{$\epsilon$}}
\newcommand{\EN}{\mbox{$\epsilon_n$}}
\newcommand{\EI}{\mbox{$\epsilon_i$}}
\newcommand{\ENI}{\mbox{$\epsilon_{ni}$}}
\newcommand{\ENX}{\mbox{$\epsilon_{nx}$}}
\newcommand{\ENY}{\mbox{$\epsilon_{ny}$}}
\newcommand{\EX}{\mbox{$\epsilon_x$}}
\newcommand{\EY}{\mbox{$\epsilon_y$}}
\newcommand{\BI}{\mbox{$\beta_i$}}
\newcommand{\BX}{\mbox{$\beta_x$}}
\newcommand{\BY}{\mbox{$\beta_y$}}
\newcommand{\SX}{\mbox{$\sigma_x$}}
\newcommand{\SY}{\mbox{$\sigma_y$}}
\newcommand{\SZ}{\mbox{$\sigma_z$}}  

\maketitle\abstracts{Photon collider based on cold and warm linear
collider technologies are compared from the point of view of
attainable luminosities, technical feasibility of laser systems and
experimental conditions. }
\section{Introduction}
   On August 20, 2004, ten day before the deadline for this
Proceedings, the International Committee for Future Accelerators
following the recommendation of International Technology
Recommendation Panel announced that International Linear Collider
(ILC) is to be realized in {\it superconducting} technology. So, all
discussions are finished. Nevertheless, even now it has sense to
summarize what is acquired and what is lost due to such decision.
I would like also to express my personal opinion about only {\it one}
linear collider (LC) in the world.

\subsection{One or two colliders?}

  All history of the mankind has proved that a monopoly in any field
  is bad, in economics monopolies are forbidden by law, why it should
  be good in science?  Competition in science is not less important
  than in economics. There are many examples.  In the case of linear
  colliders it was not necessary to create artificially  competing
  LC projects, they existed and were developed for 15--20 years by very
  strong regional collaborations and had similar readiness for the
  construction.
 
It would be wise to launch simultaneously the construction of two
   linear colliders, the superconducting, cold ``TESLA-like'' on the
   energy 0.5 $\to$ 0.8 TeV and the warm ``NLC/GLC-like'' on the energy 0.5
   $\to$ 1.5 TeV, then five year later the  CLIC on the maximum energy 3--5
   TeV. At start first two collider would have similar energy and
   study new physics together, after that TESLA  investigates more
   carefully in the sub-TeV region while NLC/GLC goes to 1.5--2 times
   higher energies. Beside the ``monopoly'' reasons, the energy region
   from 0.1 to 1.5 TeV is too large for one collider, also it takes
   long running time for investigation of all interesting energy
   points and types of collisions.

   This point of view I expressed at LCWS02 in Korea and privately
   many people told that they have similar opinion.  I believe that
   the ``competition'' argument is very important, if possible, ``global''
   projects should be avoided.

\subsection{The case of the Photon Collider}

The photon collider is based on the \EPEM\ collider. High energy
photons are produced by Compton scattering of laser photons on high
energy electrons. The expected number of interesting events is even
higher than in \EPEM\ collisions. In \GG, \GE\ collisions one can study
the same particles as in \EPEM\ collisions but in different reactions
which is very important for understanding new phenomena.  An
additional cost is mainly the cost of the laser system which is about
2-3 \% of the total LC cost (the second interaction region with the
detector is planed in any case). So, the case is obvious, paying small
incremental cost we can get new types of collision with a great physics
potential.

\section{Comparison of cold and warm photon colliders}

Below we compare two photon colliders: cold superconducting TESLA-like
collider and warm NLC/GLC-like colliders.

\subsection{Luminosities}

At sub-TeV photon colliders beam collision effects are not important
and \GG\ luminosities are just proportional to the geometric
luminosity $L_{geom}=N^2 \nu /(4\pi\sigma_x \sigma_y) \propto N^2 \nu
/\sqrt{\beta_x \beta_y \ENX \ENY}$. In TESLA the beam power $P\propto
N\nu$ is 1.65 times larger than in NLC/GLC due to 2.5 times better
efficiency of the energy transfer from the wall plug to the
beam. Moreover, TESLA can accelerate beam with larger number of
particles, which is advantageous for the photon collider ($L\propto
P\cdot N$).  The minimum value of the vertical $\beta$-function
$\beta_y \sim \sigma_z$. The minimum value of the horizontal
$\beta$-function is determined by the chromo-geometric aberrations of
the final focus system, for TESLA it is about 1.5 mm,~\cite{TESLATDR}
for NLC/GLC it is also quite similar,~\cite{NLC}$^,$~\cite{JLC} below
we assume that they are equal. Normalized emittances of electron beams
produced in damping rings are determined by synchrotron radiation,
intra-beam scattering and a tune shift due to the beam space
charge. In the TESLA case the train is much longer, therefore the
circumference $C_{\mbox{\small DR}}$ should be larger, in TESLA TDR it
is 17 km.  This causes the problem of the tune shift which is
proportional to $C_{\mbox{\small DR}} N /\sqrt{\ENX \ENY}$. At present
conference S.Mishra told us about a new approach to the damping ring
for TESLA with three times smaller circumference and by a factor of
four smaller the horizontal emittance compared to those in the TESLA
TDR for \EPEM\ collisions (for \GG\ collisions in the TESLA TDR we
assumed four times better \ENX, but it was somewhat risky number).  As
the result, the normalized transverse emittances produced by the TESLA
and NLC/GLC damping rings are almost equal. During the extraction and
acceleration the emittances are diluted somewhat, smaller in the cold
case due to smaller wake fields. The beams parameters at the
interaction point are presented in Table 1. The ratio of
geometric luminosities $L_{\mbox{{\small
TESLA}}}/L_{\mbox{\small{NLC/GLC}}} \approx 3$ in favour of the TESLA.
\begin{table}[!hbtp]
 \vspace{-5mm}
\begin{center}
\caption{Beam parameters of the cold and warm colliders}
 \vspace{3mm}
\begin{tabular}{l l l l l l }
 & $N$     & $\nu $ & $\sigma_z$ & $\beta_x/\beta_y$ &\ENX/\ENY    \\
& $10^{10}$ &  kHz   &  \MM\      &    \MM\       & $10^{-6}\,\M$ \\[2mm]
TESLA   & 2    &   14.1 & 0.3  & 1.5/0.3 & 3/0.03    \\
NLC/GLC \,\,\,\,\, & 0.75 & 23   & 0.11 & 1.5/0.11 & 3.5/0.035  
\end{tabular} \end{center} 
\end{table}
\subsection{Laser systems}
The main difference of laser systems for TESLA and NLC/GLC is
connected with the time structures of electron
trains.~\cite{telnov2002} In the TESLA: rep. rate 5 Hz , the number of
bunches in the train 2820 with the interval 337 nsec, the train
duration is 1 msec; in NLC/GLC the corresponding parameters are: 120
Hz, 192 bunches, 1.4 nsec (we assume the same structure as for \EPEM)
and 270 nsec. The train duration in the TESLA is longer than the
storage time of laser media (about 0.5-1 msec). The instantaneous
laser power during the train is about $5 \mbox{J}\times 2820 / 0.001
\sim 14$ MW for one beam, which is prohibitively large, if each laser
bunch is used only once.  For NLC/GLC one can use advantage of the
medium storage time, then the {\it effective} average power is about
$1.5\, \mbox{J}\times 192 / 0.0005 \sim 0.6$ MW, which is
acceptable. On the other hand, the large distance between electron
bunches in TESLA allows to use an external optical cavity which can
reduce the input laser power by a factor of 100, this make the laser
system much cheaper and more efficient than for the NLC/GLC. Note,
that though the external optical cavities exist in many laboratories,
but not for so short (1 psec) and powerful pulses, while for NLC/GLC
the laser system can be based on the single pass lasers developed for
fusion.

\subsection{Experimentation}

 The average number of background $\GG\ \to hadron$ events at photon
colliders is about 1--2 per bunch crossing. At TESLA each bunch
collision is seen by the detector separately.  In NLC/GLC the train is
very short and the calorimeter will integrate a whole train. However,
in Si-W calorimeter,  a very good timing resolution
is possible. The whole train is recoded but then the time for each
energy cluster can be determined with several nsec resolution.
However, in any case, the background situation at TESLA will be
better.

\subsection{The beam dump}

  High energy  photon beam at the photon colliders are very narrow,
powerful and can not be deflected by magnets. This can cause
overheating and material stress problems. The situation is more severe
for TESLA because the number of particles in one train is 40 times
larger than at NLC/GLC while the train duration is still smaller than
the thermal diffusion time. On the other hand, for the thermal stress the
characteristic times is $t_s \sim d_{heat}/v_{sound} \sim$ 1 mm/5
km/sec $\sim 2\cdot10^{-7}$ sec which integrates the whole train in
NLC/GLC and only one bunch in TESLA. A possible solution for TESLA
beam dump is considered in other my talk at this
conferences.~\cite{tel_paris2}  It is important technical
problem, but, if solved, it does not influence  photon collider
parameters.

\section{Conclusion}

The \GG\ luminosity at the cold LC can be higher by a factor of three for
present designs. The laser system is cheaper and more elegant for the
cold LC (the optical cavity), but one pass laser system for the warm
LC is more developed. The problem of the hadronic background is much
easier for the cold LC, but probably it can be solved for the warm LC as
well.  It is good that ICFA has decided to push forward the cold linear
collider. However the  era of warm linear colliders is not
finished, the CLIC project is the only feasible candidate for
the multi-TeV region.


\section*{References}
\begin{thebibliography}{99}
\bibitem{TESLATDR} B.~Badelek et al., {\it TESLA Technical Design Report,
Part VI, Ch.1.  Photon collider at TESLA}, DESY 2001-011, ECFA
2001-209, TESLA Report 2001-23, hep-ex/0108012.

\bibitem{NLC}Linear collider physics resource book for Snowmass 2001,
By American Linear Collider Working Group (T. Abe et al.), SLAC-R-570,
May 2001.  D.~Asner, J.~Gronberg, J.~Gunion, Phys.\ Rev.\ D {\bf 67},
035009 (2003).

\bibitem{JLC}
\newblock ACFA Linear Collider Working group (K.Abe et al.) KEK-REPORT-2001-11,
hep-ph/0109166.
 
\bibitem{telnov2002} V.Telnov,
\newblock  {\it Nucl. Instr.  {\rm\&}Meth.}
{\bf A494}:35, 2002.


\bibitem{tel_paris2} V.Telnov, L.Shekhtman, Conception of the beam
dump for the photon collider at TESLA, these Proceedings.
\end{thebibliography}
\end{document}